\documentclass{ws-procs9x6}

\begin{document}

\title{Fermion Masses and Neutrino Oscillations in 
$SO(10) \times SU(2)_{F}$\footnote{
\uppercase{B}ased on Plenary talk presented at 
\uppercase{PASCOS}'04, and talks
presented at \uppercase{SUSY}'04 and \uppercase{DPF}'04.}}

\author{Mu-Chun Chen}

\address{Physics Department, Brookhaven National Lab, 
Upton, NY 11973, USA \\ 
E-mail: chen@quark.phy.bnl.gov}

\author{K.T.~Mahanthappa}

\address{
Physics Department, University of Colorado, 
Boulder, CO 80309, USA\\
E-mail: ktm@pizero.colorado.edu}  
\maketitle

\abstracts{
We present in this talk 
a  model based on $SO(10) \times SU(2)_{F}$ having symmetric mass 
textures with 5 zeros constructed by us recently.  
The symmetric mass textures arising from the left-right 
symmetry breaking chain of SO(10) give rise to good predictions 
for the masses, mixing angles and CP violation measures in the 
quark and lepton sectors (including the neutrinos), 
all in agreement with the most up-to-date experimental data within $1 \; \sigma$.
Various lepton flavor violating decays 
in our model are also investigated. 
Unlike in models with lop-sided textures, our prediction for 
the decay rate of $\mu \rightarrow e \gamma$ is much suppressed and 
yet it is large enough to be probed by the next generation of experiments.
The observed baryonic asymmetry in the Universe can be accommodated 
in our model utilizing soft leptogenesis.
}

\section{Introduction}

SO(10) has long been thought to be an attractive candidate for a
grand unified theory (GUT) for a number of reasons: First of all, it unifies
all the 15 known fermions with the right-handed neutrino for each family
into one 16-dimensional spinor representation. The seesaw mechanism then
arises very naturally, 
and the small yet non-zero neutrino masses can thus be explained.
Since a complete quark-lepton symmetry is achieved, it has the promise for
explaining the pattern of fermion masses and mixing. 
Recent atmospheric neutrino oscillation data from
Super-Kamiokande indicates non-zero neutrino masses. This in turn gives very
strong support to the viability of SO(10) as a GUT group. Models based on
SO(10) combined with discrete or continuous family symmetry have been
constructed to understand the flavor problem\cite{paper1}. 
Most of the models utilize 
``lopsided'' mass textures which usually require
more parameters and therefore are less constrained. The
right-handed neutrino Majorana mass operators in most of these models are made
out of $ 16_{H} \times 16_{H}$ which breaks the R-parity at a very
high scale. The aim of this talk, based on 
Ref. [2-4], 
is to present a realistic model based on supersymmetric SO(10) combined
with SU(2) family symmetry which successfully predicts the low energy
fermion masses and mixings. Since we utilize {\it symmetric} mass textures and
$ \overline{126}$-dimensional Higgs representations for the right-handed
neutrino Majorana mass operator, our model is more constrained in addition to
having R-parity conserved. We also investigate several lepton flavor 
violating (LFV) processes in our model as well as 
soft leptogenesis\cite{paper5}. 

\section{The Model}

There are so far no fundamental understandings of the origin of flavor have been  
found.  
A less ambitious aim is to reduce the number of parameters by imposing texture 
assumptions. We concentrate on symmetric mass matrices as they are more 
predictive and can arise naturally if SO(10) is broken to the SM with 
the left-right symmetry at the intermediate scale. 
Naively one would expect that there are six texture zeros for symmetric 
quark mass matrices 
because there are six non-zero quark masses. It has been shown that this does not 
work and in order to obtain viable predictions, there can at most be 
five texture zeros. We consider the following combination for the up- and 
down-type quark Yukawa matrices with five zeros, which reads, after removing 
all the non-physical phases by rephasing various matter fields: 
\begin{eqnarray}
Y_{u, \nu_{LR}} & = & \left(
\begin{array}{ccc}
0 & 0 & a\\
0 & b e^{i\theta} & c\\
a & c & 1
\end{array}
\right) d
\label{yu}\\
Y_{d,e} & = & \left(
\begin{array}{ccc}
0 & e e^{-i\xi} & 0 \\
e e^{i\xi} & (1,-3) f & 0 \\
0 & 0 & 1
\end{array}
\right) h \quad .
\label{yd}
\end{eqnarray}

The above texture combination can be realized by utilizing an $SU(2)_{F}$ family 
symmetry. 
In order to specify the superpotential uniquely, we invoke 
$Z_{2} \times Z_{2} \times Z_{2}$ discrete symmetry. The matter fields are
\begin{displaymath} 
\psi_{a} \sim (16,2)^{-++} \quad (a=1,2), \qquad 
\psi_{3} \sim (16,1)^{+++} 
\end{displaymath}
where $a=1,2$ and the subscripts refer to family indices; the superscripts 
$+/-$ refer to $(Z_{2})^{3}$ charges. The Higgs fields which break $SO(10)$
and give rise to mass matrices upon acquiring VEV's are
\begin{eqnarray}
(10,1):\quad & T_{1}^{+++}, \quad T_{2}^{-+-},\quad
T_{3}^{--+}, \quad T_{4}^{---}, \quad T_{5}^{+--} \nonumber\\ 
(\overline{126},1):\quad & \overline{C}^{---}, \quad \overline{C}_{1}^{+++},
\quad \overline{C}_{2}^{++-}  \; .
\nonumber
\end{eqnarray}
Higgs representations $10$ and $\overline{126}$ give rise to Yukawa couplings
to the matter fields which are symmetric under the interchange of family
indices. $SO(10)$ is broken through the left-right symmetry breaking chain, and 
symmetric mass matrices and the following intra-family relations arise,
\begin{eqnarray}
M_{u} & \sim & Y^{10}_{ab} \left< 10^{+} \right> 
+ Y^{\overline{126}}_{ab} \left< \overline{126}^{+} \right>
\\
M_{d} & \sim & Y^{10}_{ab} \left< 10^{-} \right> 
+ Y^{\overline{126}}_{ab} \left< \overline{126}^{-} \right>
\\
M_{e} & \sim & Y^{10}_{ab} \left< 10^{-} \right>
- 3 Y^{\overline{126}}_{ab} \left< \overline{126}^{-} \right>
\\
M_{\nu_{LR}} & \sim & Y^{10}_{ab} \left< 10^{+} \right> 
- 3 Y^{\overline{126}}_{ab} 
\left< \overline{126}^{+}\right>
\end{eqnarray}
The $SU(2)$ family symmetry is broken in two steps and
the mass hierarchy is produced using the Froggatt-Nielsen 
mechanism:
$SU(2) \stackrel{\epsilon M}{\longrightarrow} 
U(1) \stackrel{\epsilon' M}{\longrightarrow}
nothing$
where $M$ is the UV-cutoff of the effective theory above which the family
symmetry is exact, and $\epsilon M$ and $\epsilon^{'} M$ are the VEV's
accompanying the flavon fields given by
\begin{eqnarray}
(1,2): \quad & \phi_{(1)}^{++-}, \quad \phi_{(2)}^{+-+}, \quad \Phi^{-+-}
\nonumber\\ 
(1,3): \quad & S_{(1)}^{+--}, \quad S_{(2)}^{---}, \quad
\Sigma^{++-} \; .\nonumber
\end{eqnarray}
The vacuum alignment in the flavon sector is given by 
\begin{displaymath}
\begin{array}{cccccc}
\left< \phi_{(1)} \right> & = & 
\left( \begin{array}{c} \epsilon' \\ 0 \end{array} \right),
\qquad 
&
\left< \phi_{(2)} \right> & = & 
\left( \begin{array}{c} 0 \\ \epsilon \end{array} \right)
\\
\left< S_{(1)} \right> & = & 
\left( \begin{array}{cc} 0 & \epsilon' \\ \epsilon' & 0 \end{array} \right),
\qquad
&
\left< S_{(2)} \right> & = & 
\left( \begin{array}{cc} 0 & 0  \\ 0 & \epsilon \end{array} \right)
\end{array}
\end{displaymath}
\begin{displaymath}
\begin{array}{l}
\left< \Phi \right> =  
\left( \begin{array}{c} \delta_{1} \\ \delta_{3} \end{array} \right),
\quad
\left< \Sigma \right> = 
\left( \begin{array}{cc} 0 & 0 \\ 0 & \delta_{2} \end{array} \right)
\end{array}\; .
\end{displaymath}
The various aspects of VEV's of Higgs and flavon fields are given in Ref. [2-4].

The superpotential of our model is
\begin{equation}
W = W_{Dirac} + W_{\nu_{RR}}
\end{equation}
\begin{eqnarray}
W_{Dirac}=\psi_{3}\psi_{3} T_{1}
 + \frac{1}{M} \psi_{3} \psi_{a}
\left(T_{2}\phi_{(1)}+T_{3}\phi_{(2)}\right)
\nonumber\\
+ \frac{1}{M} \psi_{a} \psi_{b} \left(T_{4} + \overline{C}\right) S_{(2)}
+ \frac{1}{M} \psi_{a} \psi_{b} T_{5} S_{(1)}
\nonumber\\
W_{\nu_{RR}}=\psi_{3} \psi_{3} \overline{C}_{1} 
+ \frac{1}{M} \psi_{3} \psi_{a} \Phi \overline{C}_{2}
+ \frac{1}{M} \psi_{a} \psi_{b} \Sigma \overline{C}_{2} \; .
\end{eqnarray}
The mass matrices then can be read from the superpotential to be
\begin{eqnarray}
M_{u,\nu_{LR}} = 
\left( \begin{array}{ccc}
0 & 0 & \left<10_{2}^{+} \right> \epsilon'\\
0 & \left<10_{4}^{+} \right> \epsilon & \left<10_{3}^{+} \right> \epsilon \\
\left<10_{2}^{+} \right> \epsilon' & \left<10_{3}^{+} \right> \epsilon &
\left<10_{1}^{+} \right>
\end{array} \right)
= 
\left( \begin{array}{ccc}
0 & 0 & r_{2} \epsilon'\\
0 & r_{4} \epsilon & \epsilon \\
r_{2} \epsilon' & \epsilon & 1
\end{array} \right) M_{U}
\end{eqnarray}
\begin{eqnarray}
M_{d,e} =  
\left(\begin{array}{ccc}
0 & \left<10_{5}^{-} \right> \epsilon' & 0 \\
\left<10_{5}^{-} \right> \epsilon' &  (1,-3)\left<\overline{126}^{-} \right>
\epsilon & 0\\ 0 & 0 & \left<10_{1}^{-} \right>
\end{array} \right)
=
\left(\begin{array}{ccc}
0 & \epsilon' & 0 \\
\epsilon' &  (1,-3) p \epsilon & 0\\
0 & 0 & 1
\end{array} \right) M_{D}
\end{eqnarray}
where
$M_{U} \equiv \left<10_{1}^{+} \right>$, 
$M_{D} \equiv \left<10_{1}^{-} \right>$, 
$r_{2} \equiv \left<10_{2}^{+} \right> / \left<10_{1}^{+} \right>$, 
$r_{4} \equiv \left<10_{4}^{+} \right> / \left<10_{1}^{+} \right>$ and
$p \equiv \left<\overline{126}^{-}\right> / \left<10_{1}^{-} \right>$.
The right-handed neutrino mass matrix is  
\begin{eqnarray}
M_{\nu_{RR}}  =   
\left( \begin{array}{ccc}
0 & 0 & \left<\overline{126}_{2}^{'0} \right> \delta_{1}\\
0 & \left<\overline{126}_{2}^{'0} \right> \delta_{2} 
& \left<\overline{126}_{2}^{'0} \right> \delta_{3} \\ 
\left<\overline{126}_{2}^{'0} \right> \delta_{1}
& \left<\overline{126}_{2}^{'0} \right> \delta_{3} &
\left<\overline{126}_{1}^{'0} \right> \end{array} \right)
=
\left( \begin{array}{ccc}
0 & 0 & \delta_{1}\\
0 & \delta_{2} & \delta_{3} \\ 
\delta_{1} & \delta_{3} & 1
\end{array} \right) M_{R}
\label{Mrr}
\end{eqnarray}
with $M_{R} \equiv \left<\overline{126}^{'0}_{1}\right>$.
Here the superscripts $+/-/0$ refer to the sign of the hypercharge. 
It is to be noted that there is a factor of $-3$ difference between the $(22)$
elements of mass matrices $M_{d}$ and $M_{e}$. This is due to the CG
coefficients associated with $\overline{126}$; as a consequence, we obtain the
phenomenologically viable Georgi-Jarlskog relation.
We then parameterize the Yukawa matrices as given in Eq.~(\ref{yu}) and 
(\ref{yd}).

We use the following as inputs at 
$M_{Z}=91.187 \; GeV$:
\begin{eqnarray}
m_{u} & = & 2.21 \; MeV (2.33^{+0.42}_{-0.45})\nonumber\\ 
m_{c} & = & 682 \; MeV (677^{+56}_{-61})\nonumber\\
m_{t} & = & 181 \; GeV (181^{+}_{-}13) \nonumber\\
m_{e} & = & 0.486 \; MeV (0.486847)\nonumber\\
m_{\mu} & = & 103 \; MeV (102.75)\nonumber\\
m_{\tau} & = & 1.74 \; GeV (1.7467) \nonumber\\
\vert V_{us} \vert & = & 0.225 (0.221-0.227)\nonumber\\
\vert V_{ub} \vert & = & 0.00368 (0.0029-0.0045)\nonumber\\
\vert V_{cb} \vert & = & 0.0392 (0.039-0.044)\nonumber
\end{eqnarray} 
where  the values extrapolated from experimental data are given inside the
parentheses. Note that the masses given above are defined in the 
modified minimal subtraction ($\overline{\mbox{MS}}$) scheme and are  
evaluated at $M_{Z}$.
These values correspond to the following set of input parameters
at the GUT scale,  $M_{GUT} = 1.03 \times 10^{16} \; GeV$, and 
$\tan\beta=10$: 
\begin{eqnarray} 
& a = 0.00250, \quad b =  3.26 \times 10^{-3}\nonumber\\
& c = 0.0346, \quad  d =  0.650\nonumber\\
& \theta  = 0.74 \nonumber\\
& e  = 4.036 \times 10^{-3}, \quad f  =  0.0195 \nonumber\\
& h =  0.06878, \quad \xi  =  -1.52 \nonumber\\
& g_{1} =  g_{2} = g_{3}  =  0.746
\nonumber
\end{eqnarray}
the one-loop renormalization group equations for the MSSM spectrum with three
right-handed neutrinos 
are solved numerically down to the effective
right-handed neutrino mass scale, $M_{R}$. At $M_{R}$, the seesaw mechanism  
is implemented. With the constraints 
$|m_{\nu_{3}}| \gg |m_{\nu_{2}}|, \; |m_{\nu_{1}}|$ and 
maximal mixing in the atmospheric sector, the up-type mass texture leads us 
to choose the following effective neutrino mass matrix
\begin{equation}\label{mll}
M_{\nu_{LL}} = \left(
\begin{array}{ccc}
0 & 0 & t\\
0 & 1 & 1+t^{n}\\
t & 1+t^{n} & 1
\end{array}
\right)\frac{d^{2}v_{u}^{2}}{M_{R}}
\end{equation}
with $n=1.15$, and from the seesaw formula we obtain
\begin{eqnarray}
\label{delta}
\delta_{1} & = & \frac{a^{2}}{r}
\\
\delta_{2} & = & \frac{b^{2}t e^{2i\theta}}
{r}
\\
\delta_{3} & = & \frac{-a(be^{i\theta}(1+t^{1.15})-c)+bct e^{i\theta}}
{r} \; ,
\end{eqnarray}
where $r=(c^{2}t+a^{2}t^{0.15}(2+t^{1.15})-2a(-1+c+ct^{1.15}))$. 
A generic feature of mass matrices of the type 
given in Eq.(\ref{mll}) is that they give rise to 
bi-large mixing pattern.  And the value of $|U_{e3}|^{2}$ is proportional to 
the ratio of $\Delta m_{\odot}^{2}$ to $\Delta m_{atm}^{2}$.

We then solve the two-loop RGE's for the MSSM spectrum 
down to the SUSY breaking scale, taken to be $m_{t}(m_{t})=176.4 \; GeV$, and
then the SM RGE's from $m_{t}(m_{t})$ to the weak scale, $M_{Z}$. 
We assume that 
$\tan \beta \equiv v_{u}/v_{d} = 10$,  with 
$v_{u}^{2} + v_{d}^{2} = (246/\sqrt{2} \; GeV) ^{2}$. At the weak scale
$M_{Z}$, the predictions for 
$\alpha_{i}
\equiv g_{i}^{2}/4\pi$ are   
\begin{displaymath} 
\alpha_{1}=0.01663,
\quad \alpha_{2}=0.03374, 
\quad \alpha_{3}=0.1242  \; .
\end{displaymath}
These values compare very well with the values extrapolated to $M_{Z}$ from the
experimental data, 
$(\alpha_{1},\alpha_{2},\alpha_{3})=
(0.01696,0.03371,0.1214 \pm 0.0031)$.
The predictions at the weak scale $M_{Z}$ for the
charged fermion masses, CKM matrix elements and strengths of CP violation, 
are summarized in Table.
The predictions for the charged fermion masses, the 
CKM matrix elements and the CP violation measures.
The predictions of our model in this {\it updated} fit are in 
good agreement with all  
experimental data within $1 \sigma$, including much improved measurements in 
B Physics that give rise to precise values for the CKM matrix elements and for 
the unitarity triangle. Note that we have 
taken the SUSY threshold correction to $m_{b}$ to be 
$-18 \%$.
\begin{table}[t]
\tbl{The predictions for the charged fermion masses, the 
CKM matrix elements and the CP violation measures.}
{\footnotesize
\begin{tabular}{@{}l c | c c l c c c l@{}}
\hline
{ } & { } & { } & { } & experimental results 
 { } &{ } & { }& { }& predictions at $M_{z}$ \\ [1ex]
{ } & { } &{ } & { }& extrapolated to $M_{Z}$ 
{ } & { }& { }& { }&{ } \\ [1ex]
\hline
$m_{s}/m_{d}$  
{ }&{ } & { }& { }& $17 \sim 25$  
{ }& { }& { }&{ } & $25$\\[1ex]
$m_{s}$ 
{ }& { }& { }& { }& $93.4^{+11.8}_{-13.0}MeV$  
{ }& { }& { }& { }& $86.0 MeV$\\[1ex]
$m_{b}$ 
{ }& { }& { }& { }& $3.00\pm 0.11GeV$  
{ }&{ } &{ } & { }& $3.03 GeV$\\[1ex]
\hline
$\vert V_{ud} \vert$
{ }& { }& { }& { }& $0.9739-0.9751$ 
{ }& { }&{ } &{ } & $0.974$\\[1ex]
$\vert V_{cd} \vert$  
{ }&{ } & { }& { }& $0.221-0.227$  
{ }& { }& { }& { }& $0.225$\\[1ex]
$\vert V_{cs} \vert$ 
{ }& { }&{ } & { }& $0.9730-0.9744$  
{ }& { }& { }& { }& $0.973$\\[1ex]
$\vert V_{td} \vert$  
{ }& { }& { }& { }& $0.0048-0.014$  
{ }& { }& { }& { }& $0.00801$\\[1ex]
$\vert V_{ts} \vert$  
{ }& { }& { }& { }& $0.037-0.043$ 
{ }& { }&{ } &{ } & $0.0386$\\[1ex]
$\vert V_{tb} \vert$  
{ }& { }& { }& { }& $0.9990-0.9992$  
{ }& { }& { }& { }& $0.999$ \\[1ex]
$J_{CP}^{q}$  
{ }&{ } & { }& { }& $(2.88 \pm 0.33) \times 10^{-5}$  
{ }& { }& { }& { }& $2.87 \times 10^{-5}$ \\[1ex]
$\sin 2\alpha$  
{ }& { }& { }&{ } & $-0.16 \pm 0.26$
{ }& { }& { }&{ } & $-0.048$ \\[1ex]
$\sin 2\beta$ 
{ }& { }& { }& { }& $0.736 \pm 0.049$ 
{ }& { }& { }& { }& $0.740$ \\[1ex]
$\gamma$  
{ }& { }& { }& { }& $60^{0} \pm 14^{0}$
{ }&  { }& { }& { }& $64^{0}$\\[1ex]
$\overline{\rho}$  
{ }& { }& { }&{ } & $0.20 \pm 0.09$
{ }& { }& { }& { }& $0.173$\\[1ex]
$\overline{\eta}$  
{ }&{ } & { }& { }& $0.33 \pm 0.05$
{ }& { }&{ } & { }& $0.366$\\[1ex]
\hline
\end{tabular}
\label{table:predict}}
\end{table}

The allowed region for the neutrino oscillation parameters 
has been reduced significantly after Neutrino 2004. 
Using the most-up-to-date 
best fit values for the mass square difference in the atmospheric sector 
$\Delta m_{atm}^{2}=2.33 \times 10^{-3} \; eV^{2}$ and 
the mass square difference for the LMA solution 
$\Delta m_{\odot}^{2}=8.14 \times 10^{-5} \; eV^{2}$ as input 
parameters, we determine 
$t = 0.344$ and $M_{R} = 6.97 \times 10^{12} GeV$, which yield   
$(\delta_{1},\delta_{2},\delta_{3}) 
= (0.00120,0.000703 e^{i \; (1.47)},0.0210 e^{i \;(0.175)})$. We obtain 
the following predictions in the neutrino sector: 
The three mass eigenvalues are give by  
\begin{equation}
(m_{\nu_{1}},m_{\nu_{2}},m_{\nu_{3}}) = (0.00262,0.00939,0.0492) \; eV \; .
\end{equation}
The prediction for the MNS matrix is
\begin{equation}
\vert U_{MNS} \vert = 
\left(
\begin{array}{ccc}
0.852 & 0.511 & 0.116\\
0.427 & 0.560 & 0.710\\
0.304 & 0.652 & 0.695
\end{array}
\right)
\end{equation}
which translates into the mixing angles in the atmospheric, 
solar and reactor sectors,
\begin{eqnarray}
\sin^{2} 2 \theta_{atm} & \equiv & \frac{
4 \vert U_{\mu \nu_{3}} \vert^{2} |U_{\tau \nu_{3}}|^{2}}
{(1-|U_{e\nu_{3}}|^2)^{2}}
= 1.00
\\
\tan^{2} \theta_{\odot} & \equiv & \frac{\vert U_{e \nu_{2}}\vert^{2}}
{|U_{e \nu_{1}}|^{2}} = 0.36
\\
\sin^{2}\theta_{13} & = & |U_{e\nu_{3}}|^{2} = 0.0134 \; .
\end{eqnarray}
The prediction of our model for the strengths of CP violation in 
the lepton sector are
\begin{eqnarray}
J_{CP}^{l} \equiv Im\{ U_{11} U_{12}^{\ast} U_{21}^{\ast} U_{22} \}
= -0.00941
\\
(\alpha_{31},\alpha_{21}) = (0.934,-1.49) \; .
\end{eqnarray}
Using the predictions for the neutrino masses, mixing angles and the 
two Majorana phases, 
$\alpha_{31}$ and $\alpha_{21}$, the matrix element for the neutrinoless double 
$\beta$ decay can be calculated and is given by   
$\vert < m > \vert = 3.1 \times 10^{-3} \; eV$, 
with the present experimental upper bound being $0.35 \; eV$.
Masses of the heavy right-handed neutrinos are
\begin{eqnarray}
M_{1} & = & 1.09 \times 10^{7} \; GeV
\label{mr1}\\
M_{2} & = & 4.53 \times 10^{9} \; GeV
\label{mr2}\\
M_{3} & = & 6.97 \times 10^{12} \; GeV \; .
\label{mr3}
\end{eqnarray} 
The prediction for the $\sin^{2}\theta_{13}$ value is $0.0134$, 
in agreement with the current bound $0.015$ at $1 \sigma$. 
Because our prediction for 
$\sin^{2}\theta_{13}$ is very close to the present sensitivity 
of the experiment, the validity of our model can be tested in 
the foreseeable future.

\section{Lepton Flavor Violating Decays and Soft Leptogenesis}

\begin{figure}[htb]
\begin{center}
\includegraphics*[angle=270,width=6.5cm]{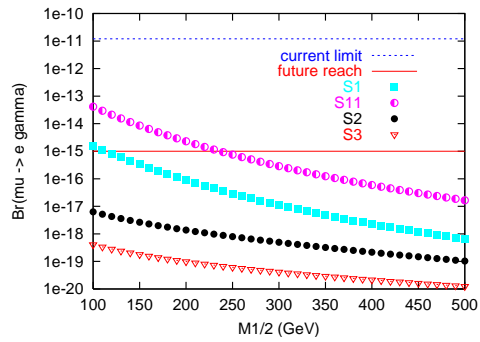}
\caption{%
The branching ratio of $\mu \rightarrow e \gamma$ as a function of 
the gaugino mass $M_{1/2}$, for various 
values of scalar masses, $m_{0}$ and $A_{0}$:
(S1): $m_{0} = A_{0} = 100 \; GeV$; 
(S11): $m_{0} = 100 \; GeV, A_{0} = 1 \; TeV$; 
(S2): $m_{0} = A_{0} = 500 \; GeV$;
(S3): $m_{0} = A_{0} = 1 \; TeV$. 
}
\label{fig1}
\end{center}
\end{figure}

Non-zero neutrino masses imply lepton flavor violation. 
If neutrino masses are 
induced by the seesaw mechanism, new Yukawa coupling involving 
the RH neutrinos can induce flavor violation. Observable decay 
rates can be obtained if the relevant scale for these LFV 
operators is the SUSY scale.
We consider LFV decays resulting from the non-vanishing 
off-diagonal matrix elements 
in the slepton mass matrix induced   
by the RG corrections  
between $M_{GUT}$ and $M_{R}$. In this case, 
the branching ratios for the decay of $\ell_{i} \rightarrow 
\ell_{j} + \gamma$ is
\begin{equation}
Br(\ell_{i} \rightarrow \ell_{j} \gamma) 
= \frac{\alpha^{3}\tan^{2} \beta}{G_{F}^{2}m_{S}^{8}} 
|\frac{-1}{8\pi} (3m_{0}^{2} + A_{0}^{2}) |^{2} 
|\sum_{k=1,2,3} (\mathcal{Y}_{\nu}^{\dagger})_{ik} 
(\mathcal{Y}_{\nu})_{kj} 
\ln (\frac{M_{GUT}}{M_{R_{k}}}) |^{2}.
\end{equation}
In our model,
$Br(\mu \rightarrow e \gamma) < Br(\tau \rightarrow e \gamma) 
< Br(\tau \rightarrow \mu \gamma)$ is predicted. 
Our predictions for the branching ratio of the decay $\mu \rightarrow e \gamma$
arising from the RG effects induced 
by neutrino Dirac Yukawa couplings as a function of the gaugino mass $M_{1/2}$ 
is given in Fig.~\ref{fig1}. 
In contrast to the predictions of models with 
lop-sided textures, in which the off-diagonal elements in 
$(23)$ sector of $M_{e}$ are of order $\mathcal{O}(1)$
leading to an enhancement in the decay branching ratio
and the need of some new mechanism to suppress the decay rate of 
$\mu \rightarrow e\gamma$, 
the predictions of our model for LFV processes, 
$\ell_{i} \rightarrow \ell_{j} \gamma$, $\mu-e$ conversion as well as 
$\mu \rightarrow 3e$, are well below the 
most stringent bounds up-to-date. Our predictions for many processes 
are nontheless within the reach 
of the next generation of LFV searches. This is especially true 
for $\mu-e$ conversion and $\mu \rightarrow e \gamma$.
More details are contained in Ref.~[5].

\begin{figure}[htb]
\begin{center}
\includegraphics*[angle=270,width=6.5cm]{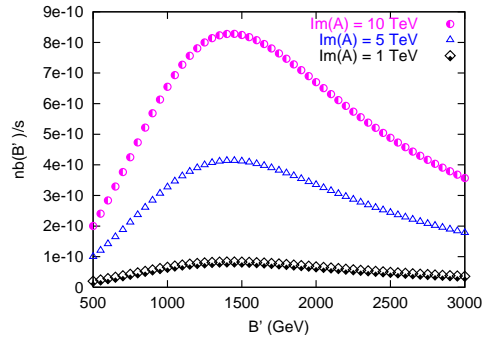}
\caption{%
The prediction for $n_{B}/s$ as a function of $B^\prime$ for 
$|Im(A)| = 10 \; TeV$, $5 \; TeV$ and 
$1 \; TeV$. 
}
\label{fig7}
\end{center}
\end{figure}

Soft leptogenesis (SFTL) utilizes the soft SUSY breaking sector,  
and the asymmetry in the lepton number is 
generated in the decay of the superpartner of the RH 
neutrinos. The lepton number asymmetry is then converted 
to the baryonic asymmetry by the sphaleron effects.
The source of CP violation in the lepton number asymmetry 
in SFTL is due to 
the CP violation in the mixing which occurs when the following 
relation 
$Im(A\Gamma_{1}/M_{1} B) \ne 0$ 
($A$ and $B$ are the tri-linear $A$-term and $B$-term)  
is satisfied. 
The total lepton number asymmetry integrated over time, $\epsilon$,  
is defined as the ratio of difference to the sum of the decay widths $\Gamma$ 
for $\widetilde{\nu}_{R_{1}}$ and $\widetilde{\nu}_{R_{1}}^{\dagger}$ 
into final states of the slepton doublet $\widetilde{L}$ and the Higgs doublet $H$, 
or the lepton doublet $L$ and the higgsino $\widetilde{H}$ or their conjugates,
\begin{eqnarray}
\epsilon & = & \frac{\sum_{f} \int_{0}^{\infty} [
\Gamma(\widetilde{\nu}_{R_{1}}, 
\widetilde{\nu}_{R_{1}}^{\dagger} \rightarrow f) - 
\Gamma(\widetilde{\nu}_{R_{1}}, \widetilde{\nu}_{R_{1}}^{\dagger} 
\rightarrow
\overline{f})]}
{\sum_{f} \int_{0}^{\infty} 
[\Gamma( \widetilde{\nu}_{R_{1}}, \widetilde{\nu}_{R_{1}}^{\dagger} 
\rightarrow f) + 
\Gamma(\widetilde{\nu}_{R_{1}}, \widetilde{\nu}_{R_{1}}^{\dagger}
\rightarrow \overline{f})]}
\end{eqnarray}
where $f$ denotes the final states $(\widetilde{L}\; H), 
\; (L \; \widetilde{H})$ and $\overline{f}$ denotes their conjugate, 
$(\widetilde{L}^{\dagger} 
\; H^{\dagger}), 
\; (\overline{L} \; \overline{\widetilde{H}})$. 
This leads to a total amount of baryon asymmetry in our model 
due to soft leptogenesis is,
\begin{equation}
\frac{n_{B}}{s} 
\simeq 
 -(1.48 \times 10^{-3}) 
\biggl( \frac{Im(A)}{M_{1}} \biggl)
\; \frac{4 \Gamma_{1} B}{\Gamma_{1}^{2} + 4 B^{2}}
 \; \delta_{B-F} \; \kappa \; .
\end{equation}
In Fig.~\ref{fig7}, we show the predictions 
for the asymmetry, $n_{B}/s$, as a function of 
$B^\prime$ for different values of $Im(A)$. 
With $B^\prime \sim 1 \; TeV$ and $Im(A) \sim 1 \; TeV$, 
sufficient baryonic asymmetry can be generated. 
More details are contained in Ref.~[5].

\section{Conclusion}

To conclude, the observed fermion mass hierarchy and mixing have been 
successfully accommodated in our model utilizing the two-step breaking in 
$SU(2)_{F}$. Due to the SO(10) and $SU(2)_{F}$ symmetries and the resulting 
symmetric mass textures, the number of parameters in the Yukawa sector 
has been significantly reduced. With $11$ parameters, our model gives rise 
to values for 12 masses, 6 mixing angles and 4 CP violating 
phases, all in agreements with available experimental data within 1 $\sigma$.
In contrast to the predictions of models with 
lop-sided textures, the predictions of our model for LFV processes, 
$\ell_{i} \rightarrow \ell_{j} \gamma$, $\mu-e$ conversion as well as 
$\mu \rightarrow 3e$, are well below the 
most stringent bounds up-to-date, and yet many of them 
are within the reach of the next generation of LFV searches. 
The observed baryonic asymmetry in the Universe can be accommodated 
in our model utilizing soft leptogenesis.

%
%
%
%
%

%
\end{document}